\documentclass[12pt]{article}
\usepackage[margin=1in]{geometry}
\usepackage{amsthm} 
\usepackage{setspace} 
\usepackage{hyperref}

\usepackage{amsmath, amssymb, amsfonts, graphicx, color, bm, float, bbm}
\usepackage{natbib}
\newcommand{\x}{\bm{x}}

\newcommand{\ABC}{\text{ABC}}

\floatstyle{ruled}
\newfloat{Algorithm}{htbp}{alg}

\begin{document}

\title{An ABC interpretation of the multiple auxiliary variable method}

\author{Dennis Prangle\footnote{Newcastle University. Email \href{mailto:dennis.prangle@newcastle.ac.uk}{\nolinkurl{dennis.prangle@newcastle.ac.uk}}} and Richard G. Everitt\footnote{University of Reading}}

\date{}

\maketitle

\begin{abstract}
We show that the auxiliary variable method \citep{Moller2006, Murray2006} for inference of Markov random fields can be viewed as an approximate Bayesian computation method for likelihood estimation.
\end{abstract}

\section{Introduction}

Markov random fields (MRFs) have densities of the form 
\begin{equation} \label{eq:MRF}
f(y|\theta) = \gamma(y|\theta) / Z(\theta),
\end{equation}
where $\gamma(y|\theta)$ can be evaluated numerically but $Z(\theta)$ cannot in a reasonable time.
This makes it challenging to perform inference.

This note considers two approaches which both use simulation from $f(y|\theta)$.
The single auxiliary variable (SAV) method \citep{Moller2006} and the multiple auxiliary variable (MAV) method \citep{Murray2006} provide unbiased likelihood estimates.
Approximate Bayesian computation \citep{Marin:2012} finds parameters which produce simulations similar to the observed data.
We will demonstrate that these two methods are in fact closely linked.

An additional challenge for inference of MRFs is that exact sampling from $f(y | \theta)$ is difficult.
It is possible to implement Markov chain Monte Carlo (MCMC) algorithms which sample from a close approximation to this distribution.
These MCMC algorithms have been used for inference through their use as a replacement for an exact sampler in SAV and MAV \citep{Caimo2011,Everitt2012} as well as ABC \citep{Grelaud:2009}.
We will use this approach and discuss it further below.

The remainder of the paper is as follows.
Section \ref{sec:background} reviews ABC and MAV methods,
and Section \ref{sec:derivation} derives our result.
Throughout the paper $y$ refers to an observed dataset,
and $x$ variables refer to simulated datasets used in inference.

\section{Background} \label{sec:background}

\subsection{Auxiliary variable methods} \label{sec:AVmethods}

The SAV method makes use of an unbiased estimate of $f(y|\theta)$, given by using the following \emph{importance sampling} (IS) estimate of $1/Z(\theta)$
\[
\widehat{\frac{1}{Z}} = \frac{q_x(x|y,\theta)}{\gamma(x|\theta)},
\]
where $q_x$ is some arbitrary (normalised) density and $x\sim f(\cdot|\theta)$. MAV extends this idea by instead using \emph{annealed IS} (AIS) \citep{Neal2001} for this estimate
\begin{equation}
\widehat{\frac{1}{Z}} = \prod_{i=2}^{a}\frac{\gamma_{i-1}(x_{i}|\theta,y)}{\gamma_{i}(x_{i}|\theta,y)}.\label{eq:1overz_mav_multiplepoints},
\end{equation}
where $f_{i}(\cdot|\theta,y)\propto \gamma_{i}(\cdot|\theta,y)$ are bridging densities between $f_{a}(\cdot|\theta,y)=f(\cdot|\theta)$ and $f_{1}(\cdot|\theta,y) = \gamma_{1}(\cdot|\theta,y) = q_{x}(\cdot|\theta,y)$, $x_a\sim f(\cdot|\theta)$ and for $2\leq i<a$, $x_i\sim K_i(\cdot |x_{i+1})$ where $K_i$ is a reversible Markov kernel with invariant distribution $f_i$.
In this description we have imposed that $\gamma_1$ is normalised in order to obtain an estimate of $1/Z$. However we note that a common choice for $q_{x}(\cdot|\theta,y)$ is $f(\cdot|\tilde{\theta})$ for some estimate $\tilde{\theta}$, in which case the normalising constant $Z(\tilde{\theta})$ is not available. In this case we obtain an estimate of $Z(\tilde{\theta})/Z(\theta)$ from Equation \eqref{eq:1overz_mav_multiplepoints}.

The SAV and MAV estimates are usually used as constituent parts of other Monte Carlo algorithms for parameter inference: in MCMC \citep{Moller2006} or IS \citep{Everitt2016}.
The estimates of $f(y|\theta)$ just described may be used here since only an unbiased estimate of the posterior up to proportionality is required \citep{Andrieu:2009}.

As noted in the introduction, the requirement of being able to draw $x_a$ exactly from $f(\cdot|\theta)$ is potentially problematic.
\cite{Caimo2011} and \cite{Everitt2012} explore the possibility of replacing this exact sampler with a long run (of $b$ iterations) of an MCMC sampler targeting $f(\cdot|\theta)$, and taking the final point. Such an approach results in biased estimates of $1/Z$, although as $b\rightarrow \infty$ this bias goes to zero. \cite{Everitt2016} observes empirically that a similar argument appears to hold when $b=0$ but $a$ is large.

\subsection{Approximate Bayesian computation}

ABC refers to a family of inference algorithms \citep[described in][]{Marin:2012} which perform an approximation to Bayesian inference when numerical evaluation of the likelihood function is intractable.
They instead use simulation from the model of interest.
The core of these algorithms is producing estimates of the likelihood $f(y|\theta)$ using some version of the following method.
Simulate a dataset $x$ from $f(\cdot|\theta)$ and return the \emph{ABC likelihood estimate}:
\[
L_\ABC = \mathbbm{1}(||y - x|| \leq \epsilon).
\]
Here $\mathbbm{1}$ represents an indicator function, $||.||$ is some distance norm, and the acceptance threshold $\epsilon$ is a tuning parameter.
The expectation of the random variable $L_\ABC$ is
\[
\int f(x | \theta) \mathbbm{1}(||y - x|| \leq \epsilon) dx.
\]
This is often referred to as the \emph{ABC likelihood}.
It is proportional to a convolution of the likelihood with a uniform density, evaluated at $y$.
For $\epsilon > 0$ this is generally an inexact approximation to the likelihood.
For discrete data it is possible to use $\epsilon=0$ in which case the ABC likelihood equals the exact likelihood, and so $L_\ABC$ is unbiased.


For MRFs empirically it is observed that, compared with competitors such as the exchange algorithm \citep{Murray2006}, ABC requires a relatively large number of simulations to yield an efficient algorithm \citep{Friel2013e}.

\section{Derivation} \label{sec:derivation}

\subsection{ABC for MRF models} \label{sec:ABCforMRFs}

Suppose that the model $f(y|\theta)$ has an intractable likelihood but can be targeted by a MCMC chain $\x = (x_1, x_2, \ldots, x_n)$.
Let $\pi$ represent densities relating to this chain.
Then $\pi_n(y|\theta) := \pi(x_n=y|\theta)$ is an approximation of $f(y|\theta)$ which can be estimated by ABC.
For now suppose that $y$ is discrete and consider the ABC likelihood estimate requiring an exact match:
simulate from $\pi(\x|\theta)$ and return $\mathbbm{1}(x_n = y)$.
We will consider an IS variation on this: simulate from $g(x|\theta)$ and return $\mathbbm{1}(x_n = y) \pi(x|\theta) / g(x|\theta)$.
Under the mild assumption that $g(x|\theta)$ has the same support as $\pi(x|\theta)$ (typically true unless $n$ is small),
both estimates have the expectation $\Pr(x_n=y|\theta)$.

This can be generalised to cover continuous data using the identity
\[
\pi_n(y|\theta) = \int_{x_n=y} \pi(\x|\theta) dx_{1:n-1},
\]
where $x_{i:j}$ represents $(x_i, x_{i+1}, \ldots, x_j)$.
An importance sampling estimate of this integral is
\begin{equation} \label{eq:ISestimate}
w = \frac{\pi(\x|\theta)}{g(x_{1:n-1}|\theta)}
\end{equation}
where $\x$ is sampled from $g(x_{1:n-1}|\theta) \delta(x_n=y)$, with $\delta$ representing a Dirac delta measure.
Then, under mild conditions on the support of $g$, $w$ is an unbiased estimate of $\pi_n(y|\theta)$.

The ideal choice of $g(x_{1:n-1}|\theta)$ is $\pi(x_{1:n-1}|x_n, \theta)$, as then $w = \pi(x_n=y | \theta)$ exactly.
This represents sampling from the Markov chain conditional on its final state being $y$.

\subsection{Equivalence to MAV}

We now show that natural choices of $\pi(\x|\theta)$ and $g(x_{1:n-1}|\theta)$ in the ABC method just outlined results in the MAV estimator \eqref{eq:1overz_mav_multiplepoints}.
Our choices are
\begin{align*}
g(x_{1:n-1}|\theta) &= \prod_{i=1}^{n-1} K_i(x_i | x_{i+1}) \\
\pi(\x|\theta) &= f_1(x_1|\theta, y) \prod_{i=1}^{n-1} K_i(x_{i+1} | x_i).
\end{align*}
Here $\pi(\x|\theta)$ defines a MCMC chain with transitions $K_i(x_{i+1} | x_i)$.
Suppose $K_i$ is as in Section \ref{sec:AVmethods} for $i \leq a$,
and for $i > a$ it is a reversible Markov kernel with invariant distribution $f(\cdot|\theta)$.
Also assume $b := n-a \to \infty$.
Then the MCMC chain ends in a long sequence of steps targeting $f(\cdot|\theta)$
so that $\lim_{n \to \infty} \pi_n(\cdot|\theta) = f(\cdot|\theta)$.
Thus the likelihood being estimated converges on the true likelihood for large $n$.
Note this is the case even for fixed $a$.

The importance density $g(x_{1:n-1}|\theta)$ specifies a reverse time MCMC chain
starting from $x_n=y$ with transitions $K_i(x_i | x_{i+1})$.
Simulating $\x$ is straightforward by sampling $x_{n-1}$, then $x_{n-2}$ and so on.
This importance density is an approximation to the ideal choice stated at the end of Section \ref{sec:ABCforMRFs}.

The resulting likelihood estimator is
\[
w = f_1(x_1|\theta, y) \prod_{i=1}^{n-1} \frac{K_i(x_{i+1} | x_i)}{K_i(x_i | x_{i+1})}.
\]
Using detailed balance gives
\[
\frac{K_i(x_{i+1} | x_i)}{K_i(x_i | x_{i+1})} = \frac{f_i(x_{i+1} | \theta, y)}{f_i(x_i | \theta, y)}
 = \frac{\gamma_i(x_{i+1} | \theta, y)}{\gamma_i(x_i | \theta, y)},
\]
so that
\[
w = f_1(x_1|\theta, y) \prod_{i=1}^{n-1} \frac{\gamma_i(x_{i+1} | \theta, y)}{\gamma_i(x_i | \theta, y)}
= \gamma(y|\theta) \prod_{i=2}^n \frac{\gamma_{i-1}(x_i | \theta, y)}{\gamma_i(x_i | \theta, y)}.
\]
This is an unbiased estimator of $\pi_n(y|\theta)$.
Hence
\[
v = \prod_{i=2}^n \frac{\gamma_{i-1}(x_i | \theta, y)}{\gamma_i(x_i | \theta, y)}
= \prod_{i=2}^a \frac{\gamma_{i-1}(x_i | \theta, y)}{\gamma_i(x_i | \theta, y)}.
\]
is an unbiased estimator of $\pi_n(y|\theta) / \gamma(y|\theta) \to 1/Z(\theta)$.
In the above we have assumed, as in Section \ref{sec:ABCforMRFs}, that $\gamma_1$ is normalised.
When this is not the case then we instead get an estimator of $Z(\tilde{\theta}) / Z(\theta)$, as for MAV methods.
Also note that in either case a valid estimator is produced for any choice of $y$.

The ABC estimate can be viewed by a two stage procedure.
First run a MCMC chain of length $b$ with any starting value, targeting $f(\cdot|\theta)$.
Let its final value be $x_a$.
Secondly run a MCMC chain $x_a, x_{a-1}, \ldots$ using kernels $K_{a-1}, K_{a-2}, \ldots$ and evaluate the estimator $v$.
This is unbiased in the limit $b \to \infty$,
so the first stage could be replaced by perfect sampling methods where these exist.

The resulting procedure is thus equivalent to that for MAV.

\section{Conclusion}
  
We have demonstrated that the MAV method can be interpreted as an ABC algorithm.
We hope this insight will be useful for the development of novel methods for MRFs.

\bibliography{MRFnote}

\end{document}